\documentclass[reprint,amsmath,amssymb,floatfix,onecolumn,superscriptaddress,aps,pre]{revtex4-2}
\usepackage{hyperref}
\usepackage{graphicx}
\usepackage{dcolumn}
\usepackage{threeparttable}
\usepackage{bm}
\usepackage{color}
\usepackage{xcolor}
\hypersetup{
    colorlinks=true,
    linkcolor=blue,
    citecolor=blue,   
   urlcolor=blue,
   }
\def\bea{\begin{eqnarray}}
\def\eea{\end{eqnarray}}
\def\beq{\begin{equation}}
\def\eeq{\end{equation}}
\def\f{\frac}

\def\t{\tau}

\def\nn{\nonumber}

\def\d{\delta}

\def\l{\lambda}
\def\L{\Lambda}

\def\d{\delta}

\begin{document}
\title{Interplay of activity and non-reciprocity in tracer dynamics: From violation of fluctuation-dissipation to giant diffusion}
\author{Subhajit Paul}
\email[]{ spaul@physics.du.ac.in} 
\thanks{These authors contributed equally to this work.}
\affiliation{Department of Physics and Astrophysics, University of Delhi, Delhi 110007, India}
\author{Manish Patel} 
\email[]{manish.patel@iopb.res.in}
\thanks{These authors contributed equally to this work.}
\affiliation{Institute of Physics, Sachivalaya Marg, Bhubaneswar-751005, India}
\affiliation{Homi Bhabha National Institute, Anushaktinagar, Mumbai-400094, India}

\author{Debasish Chaudhuri} \email[]{debc@iopb.res.in}
\affiliation{Institute of Physics, Sachivalaya Marg, Bhubaneswar-751005, India}
\affiliation{Homi Bhabha National Institute, Anushaktinagar, Mumbai-400094, India}

\date{\today}

\begin{abstract}
Non-reciprocal interactions play a key role in shaping transport in active and passive systems, giving rise to striking nonequilibrium behavior. Here, we study the dynamics of a tracer -- active or passive -- embedded in a bath of active or passive particles, coupled through non-reciprocal interactions. Starting from the microscopic stochastic dynamics of the full system, we derive an overdamped generalized Langevin equation for the tracer, incorporating a non-Markovian memory kernel that captures bath-mediated correlations. This framework allows us to compute the tracer's velocity and displacement response, formulate a generalized nonequilibrium fluctuation-dissipation relation, and determine the mean-squared displacement (MSD). We find that while the MSD becomes asymptotically diffusive, the effective diffusivity depends non-monotonically on the degree of non-reciprocity and exhibits a pronounced enhancement near a resonance set by the number of bath particles. We refer to this regime as {\it giant diffusivity}. We further show that this enhanced transport is accompanied by a strong increase in heat dissipation, revealing a direct thermodynamic cost of non-reciprocal transport. Our analytical predictions are supported by numerical simulations, including systems with short-range interactions, demonstrating the robustness of giant diffusivity and highlighting experimentally accessible signatures of non-reciprocal interactions in soft materials.

\end{abstract}


\maketitle 
\section{Introduction}
Tracer particles provide powerful probes of the microscopic and mesoscopic properties of complex media. By tracking tracer trajectories and analyzing observables such as the mean-squared displacement (MSD), one can identify distinct transport regimes, including ballistic, diffusive, subdiffusive, and superdiffusive motion, as well as dynamical crossovers that reveal relaxation mechanisms operating over different time scales~\cite{Russel1981, Zia2018, Wu2000, Kanazawa2020, Baule2023, Wang2009, Tolic2004, lutz2004, Dieterich2008, Golestanian2009, Gupta2019, Shee2020, Patel2023, paul2024sm, Patel2025}. Such tracer-based measurements form the basis of passive and active microrheology~\cite{Russel1981, Zia2018} and have been widely applied to colloidal suspensions, crowded environments, biological systems, and living matter~\cite{Wu2000, Kanazawa2020, Wang2009, Tolic2004, Benichou2018, Harris1965, Levitt1973}.
At equilibrium, tracer dynamics are typically Brownian, with diffusivity, mobility, and temperature related through the fluctuation--dissipation relation (FDR)\cite{Chaikin2012}. Away from equilibrium, however, continuous energy injection breaks detailed balance and can produce anomalous transport, non-Gaussian fluctuations, and violations of the equilibrium FDR. These effects have been extensively studied for tracers immersed in active baths composed of bacteria, active colloids, or active Brownian particles, where activity and persistence can lead to enhanced diffusion, L{\'e}vy-like statistics, and dynamical crossovers\cite{Wu2000, Kanazawa2020, Baule2023, Burkholder_Brady_2017, Burkholder2020, granek_2022}.

Tracer-based approaches are particularly valuable in biological systems, where the surrounding medium is simultaneously active, heterogeneous, and far from equilibrium. Experiments and theory have revealed anomalous diffusion in living cells~\cite{Tolic2004, barkai2012, hofling_2013, fodor2015, Otten2012}, nonequilibrium fluctuations generated by cytoskeletal activity~\cite{fodor2015, Otten2012}, and complex transport during cell migration~\cite{Dieterich2008, Podesta2017}. These observations have motivated theoretical efforts to generalize concepts such as effective temperature, the Einstein relation, and linear response to active systems~\cite{Harada_Sasa_2005, Dechant_Sasa_2020, DalCengio2019, DalCengio2021, speck2006, prost2009, Chaudhuri2014, chaudhuri2012}, as well as to understand how activity modifies viscous drag, mobility, and kinetic temperature~\cite{Loi2010, Cugliandolo2019, Mandal2019, Petrelli2020, Caprini2020}. Despite these advances, a general framework connecting microscopic system--bath interactions to tracer transport remains an active area of investigation.

A central assumption underlying most theoretical descriptions of tracer--bath dynamics is interaction reciprocity. Standard system--bath models employ reciprocal bilinear couplings, thereby enforcing action--reaction symmetry~\cite{Mori1965, zwanzig2001nonequilibrium}. Extensions to active baths generally introduce nonequilibrium effects through self-propulsion, colored noise, or modified dissipation, while retaining reciprocal interactions~\cite{Burkholder2020, Baule2023, Khali2024}. This restriction can obscure an important source of nonequilibrium behavior: the interactions themselves may violate action--reaction symmetry.

Non-reciprocal interactions have recently emerged as a generic feature of driven many-body systems rather than a special exception~\cite{Golestanian_2024, Dinelli2023}. They can arise naturally from hydrodynamic couplings, diffusive chemical fields, elastic deformations, and information-mediated interactions, and can generate qualitatively new collective behavior~\cite{You2020, Fruchart2021, Saha2020, Loos2023, Loos2020, Gupta2022, zhang_prr2023}. Such interactions have been demonstrated theoretically and experimentally in chemically active and catalytic colloids~\cite{SotoGolestanian2014, SahaGolestanianRamaswamy2014, Stark2018, NasouriGolestanian2020, Grauer2020, Mandal2024}, systems with vision-based or programmable interactions~\cite{Loos2023, Lavergne2019, Schmidt2018, barberis2016, negi2022}, and quorum-sensing active particles~\cite{Bauerle2018, Fischer2020}. They also underlie important biological and ecological interactions, including chase-and-run dynamics, predator--prey interactions, and pursuit--evasion strategies~\cite{Nagano_Maeda_2012, Sengupta2011, Xue_Goldenfeld_2016, nahin2007, isaacs1965, foreman1977, owen2008}.

While the consequences of non-reciprocity for collective phenomena such as pattern formation, traveling states, and non-reciprocal phase transitions have received considerable attention~\cite{You2020, Fruchart2021, Saha2020}, its role in single-particle transport remains much less explored. This gap is particularly relevant for tracer-based measurements, which provide direct and experimentally accessible probes of nonequilibrium environments. Determining how non-reciprocal interactions modify tracer fluctuations, response, and transport is therefore essential for interpreting microrheological measurements in active and biological systems.

In this work, we develop a generalized system-bath framework in which both the system and the bath can be either passive or active, while the reciprocity of their mutual interactions is continuously controlled by a single parameter. This construction allows us to disentangle the effects of particle activity from those arising specifically from non-reciprocal coupling. Starting from this general framework, we derive the overdamped generalized Langevin equation for the system~\cite{Mori1965, zwanzig2001nonequilibrium}, obtain its linear response function, and establish the corresponding nonequilibrium fluctuation--dissipation relation and mean-squared displacement.

Our main results can be summarized as follows. We find that the analytical expression for the system diffusivity exhibits a resonance-like divergence at a specific non-reciprocal coupling, whose value depends on the number of bath particles. Remarkably, this resonance is insensitive to whether the system and bath particles are passive or active: it is controlled solely by the degree of interaction non-reciprocity. Numerical simulations of microscopic models with physically reasonable short-ranged interactions confirm the analytical prediction and show good agreement across a broad range of parameters. The resonance, therefore, represents a generic transport consequence of non-reciprocal coupling rather than a feature tied to a particular active-particle model. Finally, we derive the heat dissipated by the system and show that it is maximized at the same non-reciprocal coupling at which the tracer diffusivity exhibits its resonant enhancement. Thus, the enhanced transport comes at a well-defined thermodynamic cost: increasing the diffusivity through non-reciprocal interactions necessarily entails increased energy dissipation.

These results identify non-reciprocity as a fundamental control parameter for tracer transport and extend conventional system--bath descriptions~\cite{Mori1965, zwanzig2001nonequilibrium} to a broader class of active, biological, and information-driven systems. In particular, the predicted resonant enhancement of diffusivity provides a direct and experimentally accessible signature of non-reciprocal interactions in nonequilibrium media.

\section{Model for tracer coupled to bath particles}
We consider a tracer at position $X_s$ coupled to a non-interacting bath of $N$ particles at positions ${x_i}$, whose dynamics are
\bea
\frac{dx_i}{dt} &=& -\lambda(x_i-X_s)+\xi_i, \nn\\
\frac{dX_s}{dt} &=& -\Lambda\sum_{i=1}^{N}(X_s-x_i)+\zeta .
\label{eom_tracer_bath}
\eea
Here, $\lambda$ and $\Lambda$ are independent coupling constants, allowing for non-reciprocal interactions. Reciprocity corresponds to $\Lambda=\lambda$, while their ratio
\begin{equation}
\chi=\frac{\Lambda}{\lambda}
\end{equation}
serves as the non-reciprocity parameter. The statistical properties of the noises $\xi_i(t)$ and $\zeta(t)$ can be chosen independently to describe passive or active bath and tracer particles. The model therefore encompasses all four combinations of passive and active tracers and baths.

\subsection*{Modeling active and passive particles and their noise correlations}
We model activity through a zero-mean colored noise $\xi_i(t)$ with exponentially decaying correlations,
\bea
\langle \xi_i(t)\xi_j(t')\rangle
= \delta_{ij}v_0^2 e^{-|t-t'|/\tau_a},
\label{eq_xixi}
\eea
where $v_0$ and $\tau_a$ denote the characteristic active speed and persistence time, respectively. This correlation captures the two-time statistics relevant to the commonly used active Brownian particle, run-and-tumble particle, and active Ornstein--Uhlenbeck particle models~\cite{paul2024sm}. For all three models, the long-time diffusivity is $\mathcal{D}=v_0^2\tau_a$; see Appendix-\ref{App_models}.

The same statistics describe an active tracer upon replacing $\xi_i$ by $\zeta$. The passive limit follows from $\tau_a\to0$ and $v_0\to\infty$ at fixed $\mathcal{D}$, giving
$\langle \xi_i(t)\xi_j(t')\rangle
\to 2 \mathcal{D}\,\delta_{ij}\delta(t-t')$,
interpreting $\mathcal{D}=\frac{k_BT}{\gamma}$.

Thus, passive and active particles are described within a unified noise framework. We denote the active speed and persistence time of the tracer and bath particles by $(v_s,\tau_s)$ and $(v_b,\tau_b)$, respectively.

\section{Dynamics of the tracer}

Integrating out the bath degrees of freedom in Eq.\eqref{eom_tracer_bath} gives the generalized Langevin equation (see Appendix-\ref{appendix_tracereqn}),
\begin{equation}
\dot X_s(t)+N\Lambda\int_{-\infty}^{t} \mathcal{G}(t-t') \dot X_s(t')\,dt' =\eta_s^{\rm eff}(t),
\label{tracer_soln3}
\end{equation}
where the Green's function $\mathcal{G}(t-t')=e^{-\lambda(t-t')}$ and the effective noise is
\begin{equation}
\eta_s^{\rm eff}(t)=\zeta(t)
+\sqrt{N}\Lambda\int_{-\infty}^{t} \mathcal{G}(t-t') \xi(t')\,dt'
\equiv\zeta(t)+\zeta_b(t).
\label{effective_noise}
\end{equation}
Thus, $\lambda^{-1}$ sets the memory time of the tracer, while $\zeta$ and $\zeta_b$ represent the intrinsic and bath-induced noise, respectively. Since the two noises are independent, $\langle\eta_s^{\rm eff}\rangle=0$ and
\begin{equation}
\langle\eta_s^{\rm eff}(t_1)\eta_s^{\rm eff}(t_2)\rangle
=\langle\zeta(t_1)\zeta(t_2)\rangle
+\langle\zeta_b(t_1)\zeta_b(t_2)\rangle .
\label{effect_noisecrl}
\end{equation}

\subsection{Characterizing the noise}
The tracer noise has the correlation
\begin{equation}
\langle\zeta(t_1)\zeta(t_2)\rangle
=v_s^2e^{-|\Delta t|/\tau_s}
=\frac{\mathcal{D}_s}{\tau_s}e^{-|\Delta t|/\tau_s},
\qquad \Delta t=t_1-t_2,
\label{tracer_noise_crl}
\end{equation}
where $\mathcal{D}_s=v_s^2\tau_s$. The passive limit is recovered by taking $\tau_s\to0$ and $v_s\to\infty$ at fixed $\mathcal{D}_s=D_s$, yielding $\langle\zeta(t_1)\zeta(t_2)\rangle\to2D_s\delta(\Delta t)$.
Using Eq.\eqref{eq_xixi} in the definition of the bath-induced noise gives
\begin{equation}
\langle\zeta_b(t_1)\zeta_b(t_2)\rangle
=N\Lambda^2v_b^2\left[
\frac{\tau_b e^{-\lambda|\Delta t|}}
{\lambda(1-\lambda^2\tau_b^2)}
+\frac{\tau_b^2e^{-|\Delta t|/\tau_b}}
{\lambda^2\tau_b^2-1}
\right].
\label{zetab_t1t2_crl}
\end{equation}
The bath-induced noise thus contains the two characteristic time scales $\lambda^{-1}$ and $\tau_b$. The apparent singularity at $\lambda\tau_b=1$ is removable; the corresponding finite limiting form is given in Appendix-\ref{app_singularity}.

For a highly persistent bath, $\tau_b\gg\lambda^{-1}$, Eq.\eqref{zetab_t1t2_crl} reduces to
\begin{equation}
\langle\zeta_b(t_1)\zeta_b(t_2)\rangle_{\rm active}
=N\left(\frac{\Lambda}{\lambda}\right)^2
\frac{\mathcal{D}_b}{\tau_b}e^{-|\Delta t|/\tau_b},
\label{zetab_et1t2_crl}
\end{equation}
where $\mathcal{D}_b=v_b^2\tau_b$. In the passive limit $\tau_b\to0$ at fixed $\mathcal{D}_b=D_b$,
\begin{equation}
\langle\zeta_b(t_1)\zeta_b(t_2)\rangle_{\rm passive}
=2ND_b\left(\frac{\Lambda}{\lambda}\right)^2
\frac{\lambda}{2}e^{-\lambda|\Delta t|}.
\label{zetab_passive_crl}
\end{equation}
Thus, for a passive bath the only remaining memory is generated by the tracer--bath coupling. In the limit $\lambda\to\infty$, this exponential kernel becomes white. The four combinations of active and passive tracer and bath particles are summarized in Table\ref{table1}.

\begin{table}[h!]
	\caption{The table presents the two-time noise correlations, $\langle \zeta(t_1) \zeta(t_2) \rangle$ and $\langle \zeta_b(t_1) \zeta_b(t_2) \rangle$, for both the tracer and the bath particles, distinguishing between active and passive cases. }\label{table1}
	\centering
	\begin{tabular}{|c|c|c|c|}
		\hline
        ~~~&~~~~&~~~~&~~\\
		~Nature of the tracer&~Nature of bath particles~~&~$\langle \zeta(t_1) \zeta(t_2)\rangle$~&~ $\langle \zeta_b(t_1) \zeta_b(t_2)\rangle $\\
		~~~&~~~~&~~~~&~~\\
		\hline
        ~~~&~~~~&~~~~&~~\\
		~Active&~~Active~~&~~$\frac{\mathcal{D}_s}{\tau_s} e^{-|t_1-t_2|/\t_s}$~& $ {{N}}\left(\frac{\Lambda}{\lambda} \right)^2 \frac{\mathcal{D}_b}{\t_b} e^{-|t_2-t_1|/\t_b}$\\
        ~~~&~~~~&~~~~&~~\\
		\hline
        ~~~&~~~~&~~~~&~~\\
		~Active&~~Passive~~&~~$\frac{\mathcal{D}_s}{\tau_s} e^{-|t_1-t_2|/\t_s}$~& $ {{N}}\left(\frac{ \Lambda}{\lambda} \right)^2 \, 2D_b \, \f{\l}{2} e^{-\l|t_2-t_1|}$\\
        ~~~&~~~~&~~~~&~~\\
        \hline
        ~~~&~~~~&~~~~&~~\\
		~Passive&~~Active~~&~~$2D_s\delta(t_1-t_2)$~&~~$ {{N}}\left(\frac{ \Lambda}{\lambda} \right)^2 \frac{\mathcal{D}_b}{\t_b} e^{-|t_2-t_1|/\t_b}$\\
        ~~~&~~~~&~~~~&\\
		\hline
        ~~~&~~~~&~~~~&\\
        ~Passive&~~Passive~~&~~$2D_s\delta(t_1-t_2)$~&~~$ {{N}}\left(\frac{ \Lambda}{\lambda} \right)^2 \, 2D_b \, \f{\l}{2} e^{-\l|t_2-t_1|}$\\
        ~~~&~~~~&~~~~&\\
        \hline
	\end{tabular}
\end{table}

\subsection{Linear response}
We introduce an external perturbation $f(t)$, measured in velocity units by absorbing the bare mobility into its definition. Equation~\eqref{tracer_soln3} then becomes, in Fourier space,
\begin{equation}
\tilde v(\omega)\left[1+N\Lambda\tilde{\mathcal G}(\omega)\right]
=\tilde\zeta(\omega)+\tilde\zeta_b(\omega)+\tilde f(\omega),
\end{equation}
where the causal memory kernel $\mathcal G(t)=e^{-\lambda t}\Theta(t)$ has transform
\begin{equation}
\tilde{\mathcal G}(\omega)=\frac{1}{\lambda+i\omega}.
\end{equation}
The velocity response function is therefore
\begin{equation}
R_v(\omega)
\equiv\frac{\langle\tilde v(\omega)\rangle}{\tilde f(\omega)}
=\frac{\lambda+i\omega}{\lambda+N\Lambda+i\omega},
\label{eq_rv}
\end{equation}
and the corresponding zero-frequency mobility is
\begin{equation}
\mu_{\rm eff}=R_v(0)=\frac{1}{1+N\Lambda/\lambda}.
\label{eq_mobility}
\end{equation}
Using $\tilde v=i\omega\tilde X_s$, the displacement response follows as
\begin{equation}
R_x(\omega)
\equiv\frac{\langle\tilde X_s(\omega)\rangle}{\tilde f(\omega)}
=\frac{R_v(\omega)}{i\omega}.
\label{eq_rx}
\end{equation}

\subsection{Generalized fluctuation-dissipation relation}
In the steady state, the velocity correlation $C_v(\tau)=\langle v(t)v(t+\tau)\rangle$ depends only on the time difference $\tau$. Since $\tilde v(\omega)=R_v(\omega)\tilde\eta_s^{\rm eff}(\omega)$ in the absence of an external force, the velocity power spectrum is 
\begin{equation}
\tilde C_v(\omega)
=\mathcal B(\omega)|R_v(\omega)|^2,
\qquad
\mathcal B(\omega)=
\frac{2\mathcal D_s}{1+\omega^2\tau_s^2}
+\frac{2N\Lambda^2\mathcal D_b}
{(\lambda^2+\omega^2)(1+\omega^2\tau_b^2)},
\label{noise_crl_amplitude}
\end{equation}
where $\mathcal B(\omega)$ is the power spectrum of the effective noise; see Appendix-\ref{appendix3}.
Using Eq.\eqref{eq_rv}, the velocity spectrum can alternatively be written as
\begin{equation}
\tilde C_v(\omega)
=\mathcal B(\omega)
\frac{\lambda^2+\omega^2}{N\Lambda\omega}\,\operatorname{Im}R_v(\omega).
\label{eq_gfdr}
\end{equation}
Equation\eqref{eq_gfdr} provides a generalized fluctuation-dissipation relation for the non-equilibrium steady state: fluctuations are related to the dissipative part of the response through a frequency-dependent factor determined by the effective noise. Unlike the equilibrium FDR, this factor is not fixed solely by temperature. The first form of Eq.\eqref{noise_crl_amplitude} remains the fundamental spectral relation and is regular in the limit $\Lambda\to0$, whereas the FDR representation in Eq.\eqref{eq_gfdr} is meaningful for $\Lambda\neq0$.

The long-time diffusivity follows from $\tilde C_v(0)=2D_{\rm eff}$,
\begin{equation}
D_{\rm eff}
=\frac{\mathcal D_s+N(\Lambda/\lambda)^2\mathcal D_b}
{(1+N\Lambda/\lambda)^2},
\label{deff_from_Cvomega}
\end{equation}
in agreement with the independent calculation from the mean-squared displacement presented later.

The corresponding displacement spectrum follows directly from $R_x(\omega)=R_v(\omega)/(i\omega)$:
\begin{equation}
\tilde C_x(\omega)=\mathcal B(\omega)|R_x(\omega)|^2.
\label{fdt_x}
\end{equation}

\begin{figure}[t!] 
\centering
\hskip -0.1cm
\includegraphics[width=10cm]{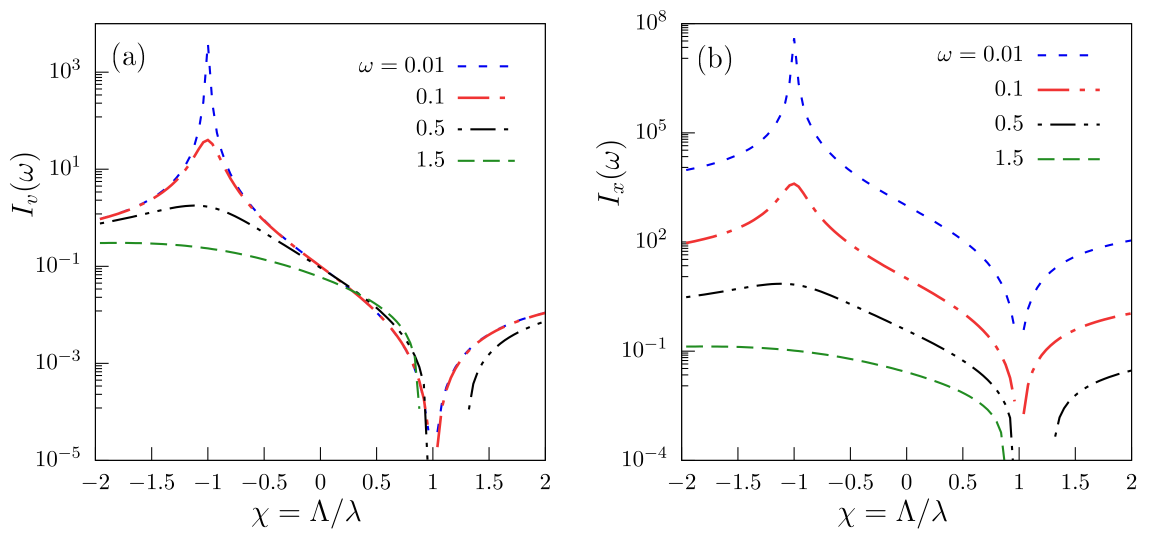}
\caption{
Plots of the nonequilibrium departures $I_v(\omega)$ in (a) and $I_x(\omega)$ in (b) as functions of the nonreciprocal parameter $\chi=\L/\l$ for different values of $\omega$. The results are obtained using Eqs.~\eqref{eq_velo_depart} and \eqref{eq_pos_depart}, respectively. The remaining parameters for both plots are set to $\mathcal{D}_s=\mathcal{D}_b=0.1$ with $\tau_s=\tau_b=0.1$, $D_{\text{eq}}=0.1$, and $N= 1$. } \label{fig_noneq_dev_fluc}
\end{figure}

\subsection{Departure from equilibrium fluctuations}
We quantify the departure from equilibrium by comparing the steady-state and equilibrium velocity spectra,
\begin{equation}
\mathcal I_v(\omega)
=\tilde C_v(\omega)-\tilde C_v^{\rm eq}(\omega),
\qquad
\tilde C_v^{\rm eq}(\omega)
=2D_{\rm eq}
\frac{\lambda^2(1+N)+\omega^2}
{\lambda^2(1+N)^2+\omega^2},
\label{eq_velo_depart}
\end{equation}
where the equilibrium reference corresponds to the reciprocal, passive limit $\chi=\Lambda/\lambda=1$, $\tau_s,\tau_b\to0$, and $\mathcal D_s=\mathcal D_b=D_{\rm eq}$. The corresponding displacement-spectrum deviation is
\begin{equation}
\mathcal I_x(\omega)
=\tilde C_x(\omega)-\tilde C_x^{\rm eq}(\omega),
\qquad
\tilde C_x^{\rm eq}(\omega)
=\frac{\tilde C_v^{\rm eq}(\omega)}{\omega^2}.
\label{eq_pos_depart}
\end{equation}

Figure~\ref{fig_noneq_dev_fluc} shows $\mathcal I_v$ and $\mathcal I_x$ as functions of $\chi = \Lambda/\lambda$ for different frequencies, using $v_s=v_b=1$ and $D_s^{\rm rot}=D_b^{\rm rot}=10$. For $N=1$, non-reciprocity strongly enhances low-frequency fluctuations near $\chi=-1$, corresponding to $\lambda+\Lambda=0$, while high-frequency modes remain comparatively insensitive. The deviations vanish in the reciprocal passive limit $\chi=1$, $\tau_s,\tau_b\to0$.

\subsection{Mean-squared displacement}
\label{sec:msd}
From Eq.\eqref{tracer_soln3}, the tracer position in Fourier space is
\begin{equation}
\tilde X_s(\omega)
=\frac{\tilde\eta_s^{\rm eff}(\omega)}
{i\omega[1+N\Lambda\tilde{\mathcal G}(\omega)]},
\qquad
\tilde{\mathcal G}(\omega)=\frac{1}{\lambda+i\omega}.
\label{tracer_fouriertransX}
\end{equation}
Using the effective-noise correlations, the MSD can be evaluated analytically; the full expression is given in Appendix-\ref{App_msd}. Its long-time behavior is diffusive,
\begin{equation}
\Delta_s(t)\simeq2 \mathcal D_s^{\rm eff} t,\qquad
\mathcal D_s^{\rm eff} =\frac{\mathcal D_s+N(\Lambda/\lambda)^2\mathcal D_b} {(1+N\Lambda/\lambda)^2},
\label{asymt_msd_tracer}
\end{equation}
in agreement with the result obtained independently from the zero-frequency velocity spectrum [Eq.~\eqref{deff_from_Cvomega}]. 

\begin{figure}[t!]
    \centering
    \includegraphics[width=0.32\linewidth]{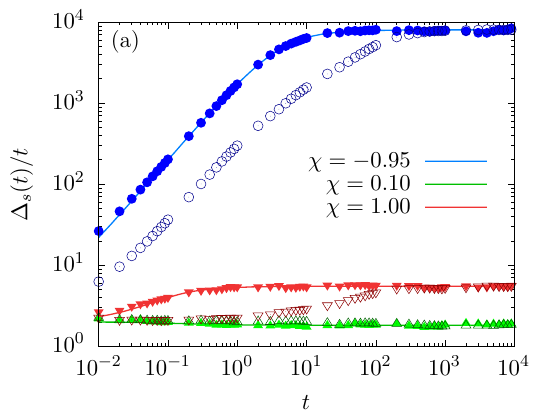}
    \includegraphics[width = 0.32\linewidth]{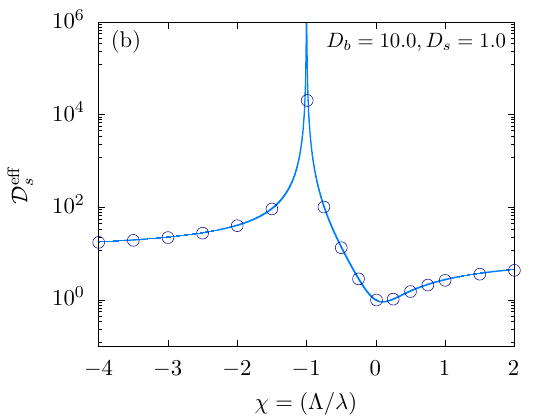}
    \includegraphics[width=0.32\linewidth]{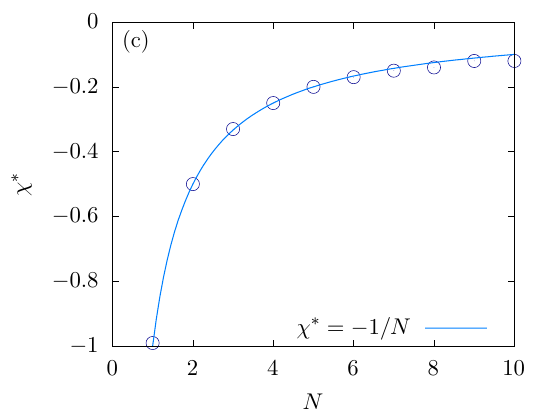}
    \caption{ (a) $\Delta_s(t)/t$ versus $t$ for several $\chi$ at $N=1$. Solid and open symbols show simulations with the harmonic interaction of Eq.\eqref{eom_tracer_bath} and the short-range interaction of Eq.\eqref{eom_harmonic_rep}, respectively; solid lines show the analytical result from Eq.\eqref{app_msd_exact}. The two models converge at late times. (b) Effective diffusivity $\mathcal D_s^{\rm eff}$ versus $\chi$ for $N=1$. Solid lines show the analytical prediction from Eq.\eqref{asymt_msd_tracer}, with a maximum near $\chi^*=-1/N$. (c) $\chi^*$ versus $N$, showing the predicted $\chi^*=-1/N$ dependence. Open symbols in (b) and (c) denote short-range simulations. For the short-range model, $L=100$, $\kappa=10$, and $r_c=20$, while $\lambda=10$ for the harmonic model; $D_s=1$ and $D_b=10$ are used for both.} 
    \label{fig:msd_deff_peak}
\end{figure}

Using the non-reciprocity parameter $\chi=\Lambda/\lambda$, the diffusivity exhibits a resonance at
\begin{equation}
\chi^*=-\frac{1}{N},
\label{eq_chi_star}
\end{equation}
where the ideal linear model predicts a divergence. Remarkably, the resonance position depends only on the non-reciprocity and the number of bath particles, and not on their active or passive nature.

We test this prediction using both the harmonic interaction in Eq.\eqref{eom_tracer_bath} and the short-range interaction described in Appendix-\ref{app:truncatedharmonic}. We integrate the corresponding equations of motion using the Euler--Maruyama scheme with time step $\d t=0.001$, taking $\lambda=10$ for the harmonic interaction and $\kappa=10$ for the short-range interaction. As shown in Fig.\ref{fig:msd_deff_peak}(a), the analytical MSD from Eq.\eqref{app_msd_exact} agrees with the harmonic-interaction simulations, while the short-range model converges to the same long-time behavior despite deviations at short times. Consequently, the effective diffusivity [Fig.\ref{fig:msd_deff_peak}(b)] exhibits a pronounced maximum near $\chi=-1$ for $N=1$. For different numbers of bath particles, the $\mathcal D_s^{\rm eff}$ versus $\chi$ curves shown in Appendix\ref{app:deff_N} reveal that the maximum shifts systematically toward $\chi^*=-1/N$, as summarized in Fig.\ref{fig:msd_deff_peak}(c). These results demonstrate that the diffusivity resonance is robust to both the microscopic form and range of the interaction. The same long-time behavior is also observed for active tracers and baths, as confirmed numerically in Appendix\ref{app:msd_ap}.

\subsection{Thermodynamic Cost of Non-Reciprocal Diffusion}
\label{sec:heat_dissipation}

The enhanced tracer diffusivity discussed in Sec.\ref{sec:msd} is accompanied by an increasing thermodynamic cost. Since the diffusivity enhancement occurs in both active and passive systems, as shown by Eq.\eqref{asymt_msd_tracer}, it is sufficient to examine the passive limit to isolate the energetic cost arising from non-reciprocity. We therefore consider the tracer and bath particles coupled to two equilibrium heat baths, with $\mathcal{D}_s=D_s$ and $\mathcal{D}_b=D_b$. 
The heat dissipated into the tracer bath is
\begin{equation}
\label{eq:dotq}
\langle\dot q\rangle
=\left\langle [\dot X_s(t)-\zeta(t)]\dot X_s(t)\right\rangle .
\end{equation}
Using Eq.\eqref{tracer_soln3}, together with the noise correlations $\langle\zeta(t)\zeta(t')\rangle=2D_s\delta(t-t')$ and $\langle\xi(t)\xi(t')\rangle=2D_b\delta(t-t')$, gives
\begin{equation}
\langle\dot q\rangle=\int_{-\infty}^{\infty}\frac{d\omega}{2\pi}\,\sigma(\omega),
\end{equation}
where
\begin{equation}
\label{eq:sigmaw}
\sigma(\omega)=
\frac{2D_bN\Lambda^2}{\lambda^2+\omega^2}
\frac{\lambda^2+N\lambda\Lambda+\omega^2}
{(\lambda+N\Lambda)^2+\omega^2}
-\frac{N\Lambda\lambda}{(\lambda+N\Lambda)^2+\omega^2}
\left[
\frac{2D_bN\Lambda^2}{\lambda^2+\omega^2}+2D_s
\right].
\end{equation}

The same combination $\lambda+N\Lambda$ that controls the tracer diffusivity also governs the heat-dissipation spectrum. For $D_s=D_b=D$, $\lambda=1$, and $\chi=\Lambda/\lambda$, Eq.~\eqref{eq:sigmaw} simplifies to
\begin{equation}
\label{eq:sigma_simple}
\sigma(\omega)=
\frac{2DN\,\chi(\chi-1)}
{(1+N\chi)^2+\omega^2}.
\end{equation}
Thus, heat dissipation is strongly enhanced as the non-reciprocity approaches the diffusivity resonance $\chi^*=-1/N$. The idealized linear model becomes singular at this point; in physical systems, nonlinearities, finite interaction ranges, and other regularizing effects will cut off the resonance. The result therefore indicates that the strong diffusivity enhancement carries a correspondingly large thermodynamic cost, rather than implying literally divergent energy dissipation.

\begin{figure} 
\centering
\hskip -0.1cm
\includegraphics[width=6.22cm]{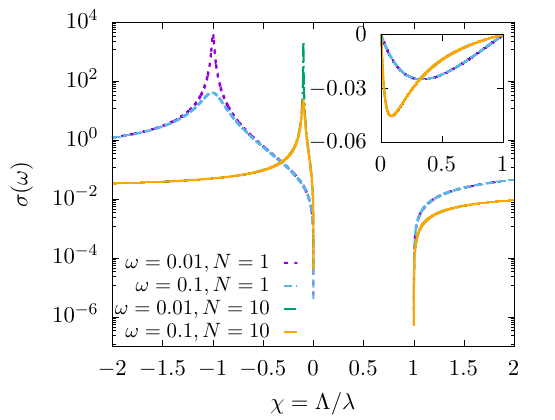}
\caption{ Heat dissipation $\sigma(\omega)$ versus the nonreciprocity $\chi=\Lambda/\lambda$ for different $\omega$ and $N$. Lines show Eq.~\eqref{eq:sigmaw} for $D_s=D_b=0.1$ and $\lambda=1$. The inset shows the corresponding results on a linear scale for $0\leq\chi\leq1$.
 } \label{fig_hp_spec}
\end{figure}

Equation~\eqref{eq:sigma_simple} also shows that $\sigma(\omega)$ vanishes at $\chi=0$ and $\chi=1$. The former corresponds to a decoupled tracer, while the latter restores reciprocal coupling and equilibrium. The negative values for $0<\chi<1$ do not violate the second law, since Eq.~\eqref{eq:dotq} measures heat transfer into the tracer bath rather than the total entropy production of the coupled system. The corresponding spectra, shown in Fig.\ref{fig_hp_spec}, exhibit strong heat-dissipation enhancement near $\chi^*=-1/N$, indicating that the giant-diffusion regime comes with a substantial thermodynamic cost and directly linking non-reciprocal transport enhancement to similar energetic dissipation.


\section{Conclusion}
\label{sec:conclusion}
We have developed a minimal and analytically tractable framework for tracer transport in active systems with non-reciprocal interactions. By integrating out a harmonically coupled bath, we obtained an overdamped generalized Langevin equation with a non-Markovian memory kernel and colored effective noise. The latter has two distinct origins: the intrinsic persistence of the particles and the memory generated by tracer--bath coupling. The framework applies equally to active and passive tracers and baths, with the passive limit recovered smoothly at fixed single-particle diffusivity. It also makes clear that, in this class of systems, reciprocity is the condition for equilibrium, whereas non-reciprocity provides a direct source of nonequilibrium behavior.

Within this framework, we derived exact response functions, the full mean-squared displacement, and the long-time effective diffusivity. We further established a generalized fluctuation--dissipation relation that provides a systematic measure of the departure from equilibrium. These results reveal a striking consequence of non-reciprocal coupling: sufficiently strong negative non-reciprocity produces a pronounced low-frequency enhancement of nonequilibrium fluctuations and, consequently, a dramatic increase of the tracer diffusivity. The location of this enhancement is controlled by the number of bath particles, with the characteristic resonance occurring near $\chi^*=-1/N$. We refer to this regime as {\it giant diffusivity}. Numerical simulations with both long-range and short-range interactions confirm the analytical predictions and show that the enhanced long-time transport persists beyond the idealized harmonic model.

Importantly, the enhancement of transport is accompanied by a thermodynamic cost. The heat-dissipation spectrum becomes strongly enhanced near the same non-reciprocal resonance, establishing a direct connection between giant diffusivity and energetic dissipation. The idealized linear theory becomes singular at the resonance, while physical nonlinearities and finite interaction ranges provide natural regularization. Thus, the resonance should be understood not as a literal divergence of physical energy consumption, but as a signature of the increasingly large energetic cost required to sustain the enhanced transport.

Taken together, our results identify non-reciprocity as a mechanism for controlling tracer transport: non-reciprocal interactions can generate exceptionally large diffusivity, but this enhancement comes at a thermodynamic cost. The dependence of the effect on coupling asymmetry and bath size provides experimentally tunable routes for controlling exploration and transport in nonequilibrium systems, with possible relevance to synthetic active matter and biological systems such as predator--prey dynamics.

\section*{Acknowledgements}
We thank Abhishek Dhar for helpful discussions. Manish Patel acknowledges the SAMKHYA high-performance computing cluster at the Institute of Physics, Bhubaneswar. Subhajit Paul acknowledges financial support from the University of Delhi through the Faculty Research Programme, Grant-IOE, Ref.~No.~IOE/2024-25/12/FRP. Debasish Chaudhuri acknowledges support from the Department of Atomic Energy, India, Grant No.~1603/2/2020/IoP/R\&D-II/15028, and an Associateship at ICTS, Bangalore, including support during ``Active Matter and Beyond'' (ICTS/ AMAB2024/11), as well as a Visiting Professorship at CY Cergy Paris Université and an Associateship at IIT Bombay.

\vskip 1cm

~~{\bf Conflicts of Interest}: There are no conflicts of interest to declare.
~~~~~
\appendix .

\section{Equivalence of active particle models}
\label{App_models}
We briefly summarize the correspondence between the commonly used active Brownian particle (ABP), run-and-tumble particle (RTP), and active Ornstein--Uhlenbeck particle (AOUP) models in one dimension~\cite{paul2024sm}. For an RTP, the active velocity is $v_0\sigma_i$, with $\sigma_i=\pm1$ switching at rate $\alpha$, giving a persistence time $\tau_a=(2\alpha)^{-1}$. For an ABP, $v_i=v_0\cos\theta_i$ with $\dot{\theta}_i=\sqrt{2D^{\rm rot}}\,\eta_i$, yielding $\tau_a=(D^{\rm rot})^{-1}$. For an AOUP, $\tau_v\dot v_i=-v_i+\sqrt{2D}\,\eta_i$, which gives $\tau_a=\tau_v$ upon identifying $v_0^2=D/\tau_v$.

At the level of two-time velocity correlations, all three models can therefore be represented by the same form,
\bea
\langle\xi_i(t)\xi_j(t')\rangle
=\delta_{ij}\,v_0^2 e^{-|t-t'|/\tau_a},
\label{eq_xixi}
\eea
with $\tau_a=(2\alpha)^{-1}$, $(D^{\rm rot})^{-1}$, and $\tau_v$ for RTP, ABP, and AOUP, respectively. In all cases, the long-time diffusivity is $\mathcal{D}=v_0^2\tau_a$. Thus, for the present analysis, the three active-particle models are equivalent at the level of their two-time correlations.

\section{Tracer dynamics}\label{appendix_tracereqn}
The tracer dynamics associated with Eq.~\eqref{eom_tracer_bath} can be obtained using the formal solution of the bath dynamics  
\begin{equation}\label{bath_solution}
	x_i(t)=x_i(0)e^{-\lambda t}+\int_{0}^{t} \Big[\xi_i(t')+\lambda X_s(t')\Big]e^{-\lambda(t- t')} dt'\,
\end{equation}
in the dynamics of $X_s$ in Eq.\eqref{eom_tracer_bath}. Using this, we obtain 
\begin{equation}
	\frac{dX_s}{dt}+N\Lambda X_s=\xi_s+\Lambda e^{-\lambda t} \sum_{i=1}^{N}x_i(0)+\Lambda \sum_{i=1}^N \Big(\int_{0}^{t} \xi_i(t') e^{-\lambda(t-t')} dt'+\lambda \int_{0}^{t}X_s(t') e^{-\lambda (t-t')}dt'\Big)\,,
\end{equation}
where the summation over $N$ describes contribution from all independent bath particles. 

The contribution from the second term vanishes either by choosing the initial conditions of the bath such that $\sum_i x_i(0)=0$, or equivalently by considering large times $t$ where the system has reached the steady state.  {Since the contributions from individual bath particles are independent, the equation for $X_s$ reduces to}
\begin{equation}\label{tracer_soln1}
\frac{dX_s}{dt}+N\Lambda X_s=\xi_s+ \sqrt{N}\Lambda \int_{0}^{t} \xi(t') e^{-\lambda(t-t')} dt'+N\Lambda \lambda\int_{0}^{t}X_s(t') e^{-\lambda(t-t')}dt'\, .
\end{equation}

\section{Bath-noise correlation at $\lambda\tau_b=1$}
\label{app_singularity}
The expression in Eq.~\eqref{zetab_t1t2_crl} contains apparent singularities when $\lambda\tau_b=1$. These singularities are removable and arise from expressing the convolution of two exponential kernels in a form containing denominators $1-\lambda^2\tau_b^2$. Taking the limit $\tau_b\to\lambda^{-1}$ gives
\begin{equation}
\left.
\langle\zeta_b(t_1)\zeta_b(t_2)\rangle
\right|_{\lambda\tau_b=1} =
\frac{N\Lambda^2v_b^2}{2\lambda^2}
\left(1+\lambda|\Delta t|\right)e^{-\lambda|\Delta t|}.
\label{eq_singularity_limit}
\end{equation}
Thus, the bath-noise correlation remains finite at $\lambda\tau_b=1$, and the apparent singularity in Eq.~\eqref{zetab_t1t2_crl} has no physical significance.

\section{Noise correlations in Fourier space}
\label{appendix3}
The bath-induced noise is related to the bath noise by
$\zeta_b(t)=\sqrt{N}\Lambda\int_{-\infty}^t
\xi(t')\mathcal{G}(t-t')\, dt'$, and hence
\begin{equation}
\tilde\zeta_b(\omega)
=\sqrt{N}\Lambda\,\tilde\xi(\omega)\tilde{\mathcal G}(\omega),
\qquad
\tilde{\mathcal G}(\omega)=\frac{1}{\lambda+i\omega}.
\label{app_zetab}
\end{equation}
Using the exponential noise correlation in Eq.~\eqref{eq_xixi}, its Fourier transform is
\begin{equation}
\langle\tilde\xi(\omega_1)\tilde\xi(\omega_2)\rangle = 
2\pi v_b^2
\left[
\frac{1}{1/\tau_b-i\omega_1}
+
\frac{1}{1/\tau_b-i\omega_2}
\right]
\delta(\omega_1+\omega_2).
\label{app_xi_corr}
\end{equation}
Therefore,
\begin{equation}
\langle\tilde\zeta_b(\omega_1)\tilde\zeta_b(\omega_2)\rangle = 
{\cal B}_1(\omega_1)\delta(\omega_1+\omega_2),
\label{corr_zetaB}
\end{equation}
where, setting $\omega_1=-\omega_2=\omega$,
\begin{equation}
{\cal B}_1(\omega) = 
\frac{4\pi N\Lambda^2{\cal D}_b}
{(\lambda^2+\omega^2)(1+\omega^2\tau_b^2)},
\qquad
{\cal D}_b=v_b^2\tau_b .
\label{B1w_form}
\end{equation}
The tracer noise has the same structure. From Eq.~\eqref{tracer_noise_crl},
\begin{equation}
\langle\tilde\zeta(\omega_1)\tilde\zeta(\omega_2)\rangle = 
{\cal B}_2(\omega)\delta(\omega_1+\omega_2),
\qquad
{\cal B}_2(\omega) = 
\frac{4\pi{\cal D}_s}{1+\omega^2\tau_s^2},
\label{corr_zeta}
\end{equation}
where ${\cal D}_s=v_s^2\tau_s$.
In the passive limit, $\tau_s,\tau_b\to0$ at fixed ${\cal D}_s=D_s$ and ${\cal D}_b=D_b$, these expressions reduce to the corresponding white-noise spectra. In particular, for the equilibrium case $\Lambda=\lambda$ and $D_s=D_b=D_{\rm eq}$,
\begin{equation}
{\cal B}^{\rm eq}(\omega)
={\cal B}_1^{\rm eq}(\omega)+{\cal B}_2^{\rm eq}(\omega)
= 4\pi D_{\rm eq}
\left[
1+\frac{N\lambda^2}{\lambda^2+\omega^2}
\right].
\label{app_Beq}
\end{equation}
These noise spectra ${\cal B}_1(\omega)$ and ${\cal B}_2(\omega) $  are used in Eqs.~\eqref{noise_crl_amplitude} and \eqref{app_Cv}.

\section{Calculation of the mean-squared displacement}
\label{App_msd}

From Eq.~\eqref{tracer_soln3}, the tracer velocity in Fourier space is
\begin{equation}
\tilde v(\omega)
=
R_v(\omega)\tilde{\eta}_s^{\rm eff}(\omega),
\qquad
R_v(\omega)
=
\frac{\lambda+i\omega}{c+i\omega},
\qquad
c\equiv\lambda+N\Lambda .
\label{app_vomega}
\end{equation}
Using the effective-noise spectrum obtained in Appendix~\ref{appendix3},
\begin{equation}
{\cal B}(\omega)
=
\frac{2{\cal D}_s}{1+\omega^2\tau_s^2}
+
\frac{2N\Lambda^2{\cal D}_b}
{(\lambda^2+\omega^2)(1+\omega^2\tau_b^2)},
\label{app_Bomega}
\end{equation}
the velocity power spectrum is
\begin{equation}
\tilde C_v(\omega)
=
{\cal B}(\omega)\left|R_v(\omega)\right|^2.
\label{app_Cv}
\end{equation}
The MSD follows from
$\Delta_s(t)=2\int_0^t(t-\tau)C_v(\tau)\,d\tau$.
Evaluating the resulting frequency integrals gives
\begin{align}
\Delta_s(t)
={}&
\frac{2{\cal D}_s}{1-c^2\tau_s^2}
\bigg[
\frac{\lambda^2}{c^3}
\left(ct-1+e^{-ct}\right)
-\lambda^2\tau_s^3
\left(\frac{t}{\tau_s}-1+e^{-t/\tau_s}\right)
\nonumber\\
&\hspace{2.8cm}
-\frac{e^{-ct}-1}{c}
+\tau_s\left(e^{-t/\tau_s}-1\right)
\bigg]
\nonumber\\
&+
\frac{2N\Lambda^2{\cal D}_b}{1-c^2\tau_b^2}
\bigg[
\frac{1}{c^3}
\left(ct-1+e^{-ct}\right)
-\tau_b^3
\left(\frac{t}{\tau_b}-1+e^{-t/\tau_b}\right)
\bigg].
\label{app_msd_exact}
\end{align}

The apparent singularities at $c\tau_s=1$ and $c\tau_b=1$ are
removable and arise only from the partial-fraction decomposition used
to evaluate the frequency integrals. The corresponding limits of
Eq.~\eqref{app_msd_exact} are finite.

At short times, Eq.~\eqref{app_msd_exact} gives ballistic motion,
$\Delta_s(t)\propto t^2$, whereas at long times it becomes diffusive,
\begin{equation}
\Delta_s(t)\simeq2D_{\rm eff}t,
\qquad
D_{\rm eff}
=
\frac{{\cal D}_s
+N(\Lambda/\lambda)^2{\cal D}_b}
{\left(1+N\Lambda/\lambda\right)^2},
\label{app_long_msd}
\end{equation}
which is the result quoted in Eq.~\eqref{asymt_msd_tracer}. Since
$D_{\rm eff}$ depends on the tracer and bath dynamics only through
their effective diffusivities ${\cal D}_s$ and ${\cal D}_b$, the
location of the diffusivity resonance is independent of whether the
tracer and bath particles are active or passive.

\begin{figure}[th!]
    \centering
    \includegraphics[width=8cm]{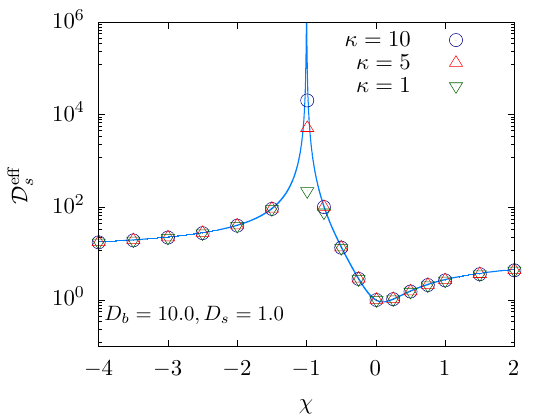}
    \caption{ Tracer diffusivity $\mathcal{D}_s^{\rm eff}$ for different values of the interaction strength $\kappa$. The parameter values are chosen as $D_s = 1$, $D_b = 10$, $L = 100$, and $r_c = 20$. The line through the simulation data points is the plot of the analytical expression of effective diffusivity obtained in Eq.~(\ref{asymt_msd_tracer}).}
    \label{fig:deff_kappa_var}
\end{figure}

\section{Short-range harmonic interaction}
\label{app:truncatedharmonic}
To test the robustness of the analytical predictions beyond the linear long-range coupling, we consider a short-range interaction generated by a shifted harmonic potential truncated at a cutoff distance $r_c$,
\begin{align}
\dot x_i
&=
\begin{cases}
-\kappa\left(|x_{is}|-r_c\right)\operatorname{sgn}(x_{is})+\xi_i,
& |x_{is}|\leq r_c,\\[4pt]
\xi_i,
& |x_{is}|>r_c,
\end{cases}
\nn\\[8pt]
\dot X_s
&=
\sum_{i=1}^{N}
\begin{cases}
\chi\, \kappa\left(|x_{is}|-r_c\right)\operatorname{sgn}(x_{is}) +\zeta ,
& |x_{is}|\leq r_c,\\[4pt]
\zeta ,
& |x_{is}|>r_c,
\end{cases}.
\label{eom_harmonic_rep}
\end{align}
with $\chi=\Lambda/\lambda$ controlling the non-reciprocity. Simulations with periodic boundary conditions are used to obtain the results shown in Fig.~\ref{fig:msd_deff_peak}.

The parameters $\kappa$ and $r_c$ control the effective interaction strength. Near the resonance, sufficiently strong interactions are required to prevent particle crossing and maintain the intended chase--run dynamics. Figure~\ref{fig:deff_kappa_var} shows that for $\kappa=1$ and $5$, particle crossing produces deviations from the analytical prediction near $\chi^*$, whereas for $\kappa\ge10$ the numerical results converge to the analytical behavior. Thus, the short-range model reproduces the predicted resonance when the interaction is sufficiently strong.

\begin{figure}[th!]
    \centering
    \includegraphics[width=16 cm]{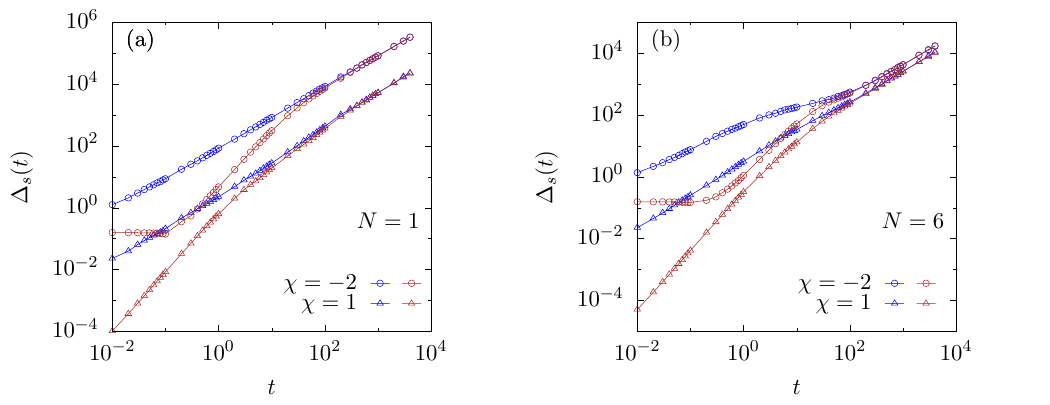}
    \caption{Comparison of the MSD for active and passive systems with short-range harmonic interactions. Blue and brown symbols denote passive and active systems, respectively. The simulations use $L=100$, $\kappa=10$, and $r_c=20$, with $D_s=1$, $D_b=10$ for the passive system and ${\mathcal D}_s=1$, ${\mathcal D}_b=10$ for the active system, corresponding to $v_s=1$, $\tau_s=1$, $v_b=1$, and $\tau_b=10$.}
    \label{fig:msd_ap}
\end{figure}

\section{Comparison of MSD in active and passive systems}
\label{app:msd_ap}
The late-time diffusivity of an active system can be mapped onto that of a thermal system. In Fig.~\ref{fig:msd_ap}, we demonstrate this equivalence by comparing a passive system with $D_s=1.0$ and $D_b=10$ to an active system with $v_s=1$, $\tau_s=1$ and $v_b=1$, $\tau_b=10$, corresponding to ${\cal D}_s=v_s^2\tau_s=1$ and ${\cal D}_b=v_b^2\tau_b=10$. For both $N=1$ and $N=6$ bath particles, the MSDs exhibit identical late-time diffusivities, despite their initial deviation due to active persistence. Thus, under the appropriate active–passive mapping, the effective diffusion coefficient is invariant.

\begin{figure}[ht!]
    \centering
    \includegraphics[width=8 cm]{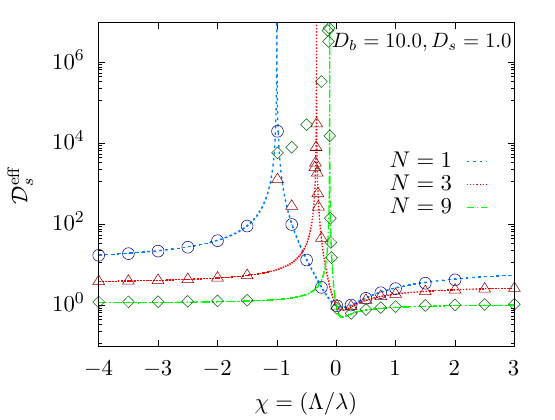}
    \caption{Tracer diffusivity ${\mathcal D}^{\rm eff}_s$ versus $\chi$ for different $N$. Lines show the analytical prediction from Eq.~(\ref{asymt_msd_tracer}), and symbols show simulations with short-range harmonic interactions. Simulations use periodic boundary conditions with $L=100$, $\kappa=20$, $r_c=20$, $\lambda=1$, $D_s=1$, and $D_b=10$.
    }
    \label{fig:deff_as_N}
\end{figure}

\section{Tracer diffusivity for varying number of bath particles}
\label{app:deff_N}
We consider a tracer with diffusivity ${\mathcal D}^{\rm eff}_s$ interacting non-reciprocally with $N$ bath particles of diffusivity $D_b$. The late-time diffusivity, Eq.(\ref{asymt_msd_tracer}), predicts a maximum at $\chi^*=-1/N$. Figure~\ref{fig:deff_as_N} compares this prediction with simulations of the short-range harmonic interaction for $L=100$, $\kappa=20$, and $r_c=20$. The simulations agree with the analytical result and capture the maximum in diffusivity near the predicted $\chi^*$. Figure~\ref{fig:msd_deff_peak}(c) shows the corresponding $\chi^*$ for different $N$, confirming the $\chi^*=-1/N$ scaling.


%
\end{document}